\documentclass{aa}
\usepackage{amssymb}
\usepackage{amsmath}
\usepackage{txfonts}
\usepackage{graphicx}
\usepackage{natbib}

\begin{document}

\title{Long term variability of Cygnus~X-1}
\subtitle{III. Radio--X-ray correlations}

\author{
  T.~Gleissner\inst{1} \and 
  J.~Wilms\inst{1}\fnmsep%
  \thanks{\emph{Present address:} Department of Physics, University of
  Warwick, Coventry CV4 7AL, UK}
  \and 
  G.~G.~Pooley\inst{2} \and
  M.~A.~Nowak\inst{3} \and 
  K.~Pottschmidt\inst{4,5} \and 
  S.~Markoff\inst{3}\fnmsep%
  \thanks{NSF Astronomy \& Astrophysics Postdoctoral Fellow}
  \and \\
  S.~Heinz\inst{3}\fnmsep%
  \thanks{\textsl{Chandra} Postdoctoral Fellow}
  \and
  M.~Klein-Wolt\inst{6} \and 
  R.~P.~Fender\inst{6} \and 
  R.~Staubert\inst{1}} 
\institute{Institut f\"ur Astronomie und Astrophysik -- Abt.~Astronomie, 
  Universit\"at T\"ubingen, Sand 1, 72076 T\"ubingen, Germany 
\and Mullard Radio Astronomy Observatory, Cavendish Laboratory, Madingley
  Road, Cambridge CB3 0HE, UK 
\and  Massachusetts Institute of Technology, Center for Space Research, 77
  Massachusetts Ave., Cambridge MA 02139, USA
\and Max-Planck-Institut f\"ur extraterrestrische
  Physik, Postfach 1312, 85748 Garching, Germany 
\and INTEGRAL Science Data Centre, Chemin d'\'Ecogia 16, 1290 Versoix,
  Switzerland  
\and Astronomical Institute ``Anton Pannekoek'', University of Amsterdam,
  and Center for High Energy Astrophysics, Kruislaan~403, 1098 SJ,
  Amsterdam, the Netherlands}

\offprints{T.~Gleissner,\\
\email{gleiss@astro.uni-tuebingen.de}}
\titlerunning{Long term variability of Cygnus~X-1: III.}
\authorrunning{T.~Gleissner et al.}
\date{Received $<$date$>$ / Accepted $<$date$>$ }

\abstract{Long time scale radio--X-ray correlations in black holes during
  the hard state have been found in many sources and there seems to emerge
  a universal underlying relationship which quantitatively describes this
  behavior. Although it would appear only natural to detect short term
  emission patterns in the X-ray and -- with a certain time lag -- in the
  radio, there has been little evidence for this up to now. The most
  prominent source for radio--X-ray correlations on short time scales
  (minutes) so far remains GRS~1915+105 where a single mass ejection could
  be detected successively in the X-ray, IR, and radio wavebands. We
  analyze a database of more than 4 years of simultaneous radio--X-ray data
  for Cygnus~X-1 from the Ryle Telescope and \textsl{RXTE PCA/HEXTE}. We
  confirm the existence of a radio--X-ray correlation on long time scales,
  especially at hard energies. We show that apparent correlations on short
  time scales in the lightcurves of Cygnus~X-1 are most likely the
  coincidental outcome of white noise statistics. Interpreting this result
  as a breakdown of radio--X-ray correlations on shorter time scales, this
  sets a limit to the speed of the jet.  } \maketitle

\section{Introduction}\label{sec:intro}

It has become generally accepted that our Galaxy harbors a class of X-ray
binaries with relativistic outflows. These objects show an intriguing
similarity to quasars and, more generally, to AGN except for the fact that
their masses are lower by a factor between $10^6$ and $10^9$. Thus they
have been referred to as `microquasars' (MQs) \citep{mirabel:98a}.  One of
the key properties of MQs is the ejection of matter at relativistic speeds
in a jet from which non-thermal radio emission is detected, which is
thought to be synchrotron radiation from relativistic electrons in the jet
\citep[for reviews see, e.g.,][]{hjellming:95a,mirabel:99a,fender:02a}. The
radio emission of X-ray binaries as well as jets seems to be related to the
hard state whereas the soft state shows a considerable reduction of radio
emission or even its quenching below detectability
\citep{fender:99,corbel:00a}.

Radio--X-ray correlations on time scales of days and months have been found
for many BHs in the hard state and even seem to follow a universal
relationship \citep{gallo:03a,corbel:03a,merloni:03a,falcke:04a}. When
proceeding to shorter time scales it becomes harder to establish a
corresponding correlation as the increasing scatter blurs the picture. A
prime example for radio--X-ray correlations on short time scales is the MQ
GRS~1915+105, where correlated radio/IR and X-ray variations have been
reported by, e.g., \citet{pooley:97}, \citet{fender:97a},
\citet{fender:00b}, and \citet{mirabel:98a} who interpreted the system in
terms of a synchrotron bubble model \citep{laan:62,hjellming:88}. The
evidence for radio--X-ray correlations on short time scales in other MQs,
such as GX~339$-$4, is much less clear \citep{corbel:00a, corbel:03a}.

The presence of correlations between the radio emission and the X-rays
suggests a direct connection between the physical processes producing the
radiation.  X-rays from accreting black holes (BHs) are usually thought to
originate from the innermost part of the accretion disk and from a
so-called accretion disk corona consisting of a high-temperature medium
suitable to upscatter soft photons to higher energies
\citep{shakura:73,sunyaev:80}. In such a scenario it is likely that a
varying mass accretion rate, $\dot{M}$, also results in a time variable
X-ray flux. Although the underlying process of mass ejections from a BH via
a relativistic outflow is still being discussed, it is beyond doubt that
part of the accreted mass stream has to be redirected and expelled as a jet
\citep{fender:02a}. Thus the variability of $\dot{M}$ should be partly
reflected in the outflow stream and in the radio flux originating from
there as well. Consequently it may be possible to detect radio--X-ray
correlations on the same time scales as X-ray variability. In the context
of X-rays from the accretion flow itself, the observed long term
radio--X-ray correlation arises naturally only for radiatively inefficient
flows \citep{heinz:03,merloni:03a}. Alternatively, in recent years it was
proposed that part of the X-ray emission is due to synchrotron processes in
the jet \citep{markoff:01a,markoff:03a}. This controversial model is
supported by its ability to describe broadband radio through X-ray
simultaneous spectra self-consistently, and also provides a natural
explanation for the long term radio--X-ray correlations seen in some BHs.

Cyg~X-1/HDE~226868 is one of the first high-mass X-ray binaries that were
detected and the binary's compact component is believed to be a BH.
The X-ray luminosity is high, making it one of the brightest X-ray sources
in the sky; furthermore it displays strong variability on time scales from
milliseconds to years. In the hard state it is also detectable in the radio
with a mean level of $\sim$14\,mJy at cm wavelengths while it
shows little or no radio emission in the soft state \citep[][and references
therein]{pooley:01a,brocksopp:99a}. The assumption that Cyg~X-1 ranks among
the MQs has been firmed up by the detection of a relativistic jet
\citep{stirling:98,spencer:01a,stirling:01a}.

Monitoring of Cyg~X-1 simultaneously in radio and X-ray bands has shown
that on long time scales of days to months or more these two kinds of
emission are loosely correlated \citep{pooley:99a,brocksopp:99a}, although
this correlation is partly smeared out due to contamination from the bright
companion star (e.g., the stellar wind of the Cyg~X-1 companion affects the
radio dispersion). In this paper we discuss the possibilities for
radio--X-ray correlations on short time scales (= minutes to maximally 10
hours) in Cyg~X-1. We perform cross-correlations between X-ray and radio
lightcurves with a relative shift of about $\pm$10\,h, which seems to be a
sensible choice as the time scale for mass ejections to travel from the
vicinity of the BH, i.e., the X-ray producing region, to the outer areas of
the jet, i.e., the radio emission region, is in the same time
range. We also analyze radio--X-ray correlations on long time scales
(i.e., days to months) in Cyg~X-1, where we compare flux--flux plots for soft
X-ray and radio, and for hard X-ray and radio, respectively.

Section~\ref{sec:data} describes the observations and the data analysis. We
test the sensitivity of our procedures to radio--X-ray correlations in
Sect.~\ref{sec:resultscyg}, where we apply our methods to GRS~1915+105, and
then present our results for Cyg~X-1. We finally discuss these results in
Sect.~\ref{sec:discussion}. Preliminary results had been published by
\citet{gleissner:04a}.

\section{Observations and data analysis}\label{sec:data}

\begin{figure*}
\centering
   \includegraphics[width=17cm]{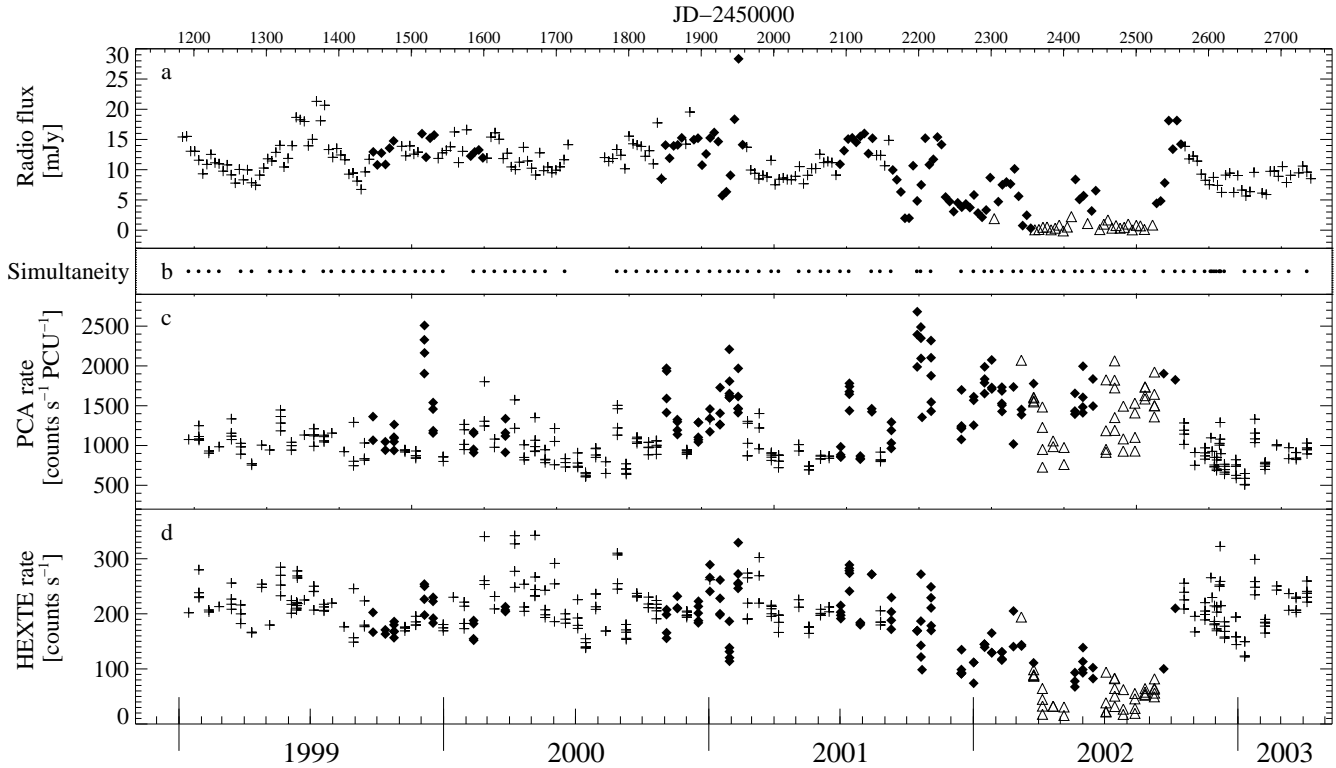}
   \caption{Cyg~X-1: \textbf{a)} Mean Ryle Telescope radio flux, rebinned to a
     resolution of 5.6\,days, the orbital period of Cyg~X-1.  \textbf{b)}
     Times of simultaneous radio--X-ray observations with Ryle Telescope
     and \textsl{RXTE}.  \textbf{c)} Mean \textsl{RXTE} \textsl{PCA} count
     rate. \textbf{d)} Mean \textsl{RXTE} \textsl{HEXTE} count rate. Hard
     states are indicated by crosses, soft states by triangles, and failed
     state transitions (FST) by filled diamonds (see text for state
     definition).}
   \label{fig:overview}
\end{figure*}

For the studies presented here, we used X-ray data from the Rossi X-ray
Timing Explorer (\textsl{RXTE}) and radio data from the Ryle Telescope at
the Mullard Radio Astronomy Observatory, Cambridge, UK. The radio data were
taken at 15\,\mbox{GHz} ($\lambda=2\,\text{cm}$) and the resulting
lightcurves have a time resolution of 8\,sec. Every $\sim$30\,min there is
a data gap of $\sim$3.5\,\mbox{min} in the radio lightcurves due to
calibration. The basic parameters of the radio telescope are described by
\citet{pooley:97}. In Fig.~\ref{fig:overview}a the radio flux of Cyg~X-1 is
displayed, binned to a resolution of 5.6\,days, the orbital period of
Cyg~X-1 \citep{brocksopp:99a}. All flux points are assigned to symbols
according to their state, following a state definition which is based on
the spectral photon index and the time lag between soft and hard photons
\citep{benlloch:03,benlloch:04}, amended by the criterion that a soft state
must have a radio flux $\leq 3$\,mJy. This definition distinguishes three
states: The canonical hard and soft states, as well as the failed state
transitions (FST) \citep{pottschmidt:00a}, which comprise all observations
which do not fall into either of the other two states.

A good part of the X-ray data of Cyg~X-1 analyzed in this work are the
lightcurves that were already used in Paper~I \citep{pottschmidt:03} and
Paper~II \citep{gleissner:04b} of this series. We therefore only give a
brief summary of the \textsl{RXTE} campaign and the data extraction issues
and refer to Paper~I for the details. The analyzed data span the time from
1999 until the beginning of 2003, covering our \textsl{RXTE} monitoring
programs P40099 (1999), P50110 (2000--2002), and P60090 (2002--2003)
(Fig.~\ref{fig:overview}c and d). Observations of a duration of 10\,ks were
performed every 2 weeks. After screening the data for episodes of increased
background, we extracted \textsl{Standard2f} lightcurves with a resolution
of 16\,s from the \textsl{RXTE} Proportional Counter Array
\citep[\textsl{PCA};][]{jahoda:96} data using the standard \textsl{RXTE}
data analysis software, \texttt{HEASOFT}, Version~5.2.  Also high energy
data from the \textsl{RXTE} High Energy X-ray Timing Experiment
\citep[\textsl{HEXTE};][]{rothschild:98} with a time resolution of 1\,s
were extracted. The \textsl{PCA} is mainly sensitive to photons in the
energy range from $\sim$2\,keV to $\sim$15\,keV, the \textsl{HEXTE} from
$\sim$15\,keV to $\sim$250\,keV (channels 15--255).

\begin{figure}
  \resizebox{\hsize}{!}{\includegraphics{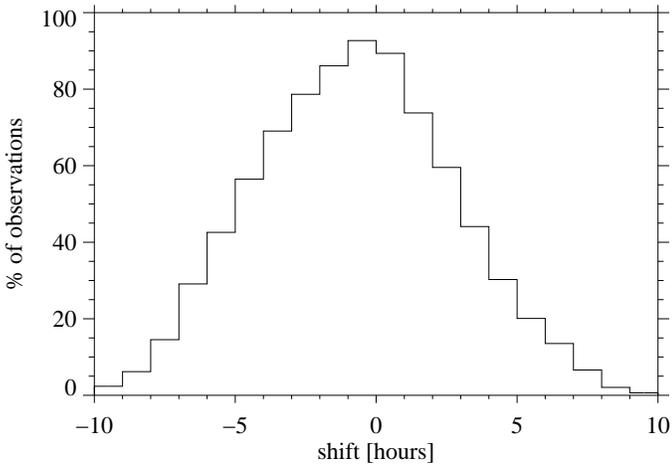}}
  \caption{Cyg~X-1: Histogram of the relative shift between the radio and X-ray
    lightcurves that is covered by the used cross-correlation
    calculations for the 301 \textsl{PCA} lightcurve segments.  There
    are only a few cases where the maximum relative shift extends up
    to 10\,h, limiting the significance of our analysis to time scales
    of less than $\sim$5\,h.}
  \label{fig:shift}
\end{figure}

In our scheduling of the monitoring of Cyg~X-1 in radio and X-ray bands, we
strived for strictly simultaneous observation times. The observations which are
analyzed here are marked with the dots in Fig.~\ref{fig:overview}b. Due to
the \textsl{RXTE} orbit, the X-ray data of each observation are composed of
several short lightcurve segments of a typical duration of $\sim$0.5\,h,
while the radio data generally consists of a longer continuous lightcurve
of several hours.  

For the analysis, the radio and X-ray lightcurves of all simultaneous
observations were rebinned to a time resolution of 32\,s. Only lightcurves
with a length of at least 15\,min were considered. We then applied the
\citet{savitzky:64} smoothing filter, i.e., a least squares polynomial fit
which can be used to smooth a noisy signal \citep{press:92}. After this
smoothing, every single continuous X-ray lightcurve segment and radio
lightcurve within a time lag of $\pm$10\,h to each other was
cross-correlated. The cross-correlation coefficient is calculated only when
the overlap of the radio and X-ray lightcurves is more than 15\,minutes.
Thus we find the maximum cross-correlation coefficient, MCC, and the
corresponding time lag for each radio and X-ray lightcurve pair.  Depending
on the data sampling and the length of the lightcurves, the maximum
possible relative shift between the radio and X-ray lightcurves is about
$\pm$10\,h, in most cases, however, the maximum shift is much less.
Fig.~\ref{fig:shift} gives the distribution of the relative shifts between
the radio and \textsl{PCA} X-ray lightcurves which are covered by our
calculations for Cygnus~X-1.  A negative shift means that the X-rays
precede the radio. As the bulk of observations covers only a relative shift
of $\sim$5\,h, our analysis is significant on time scales from minutes to
$\sim$5\,h.

Due to the \textsl{PCA} dead time after processing an event, the detected
X-ray count rate is diminished with respect to the actual count rate. To
test whether a correction for this dead time effect was necessary we
followed the description given by the \textsl{RXTE} Guest Observer Facility
(GOF), using the Very Large Event (VLE) deadtimes as given by
\citet{jernigan:00a}.  We verified that the influence of deadtime
correction of the \textsl{PCA} lightcurves on the resulting MCC and time
lag is small: when calculating the cross-correlation with and without
deadtime correction, the results of only 8 observations out of 87 had
significantly changed.  Due to the failure of the propane layer since 2000
May 12, the measured VLE rate is influenced by a higher event rate in PCU
number 0.  Since the \textsl{PCA} data modes employed by us do not contain
the PCU number, it is not possible to ignore data from this PCU in the data
analysis, and it is not possible to compute a reliable deadtime correction.
Due to this fact, we have chosen not to apply the deadtime correction in
our X-ray lightcurves.

\section{Radio--X-Ray Correlations in Cyg~X-1}\label{sec:resultscyg}

\subsection{Short term correlations}\label{sec:Short-term}

\begin{figure}
\resizebox{\hsize}{!}{\includegraphics{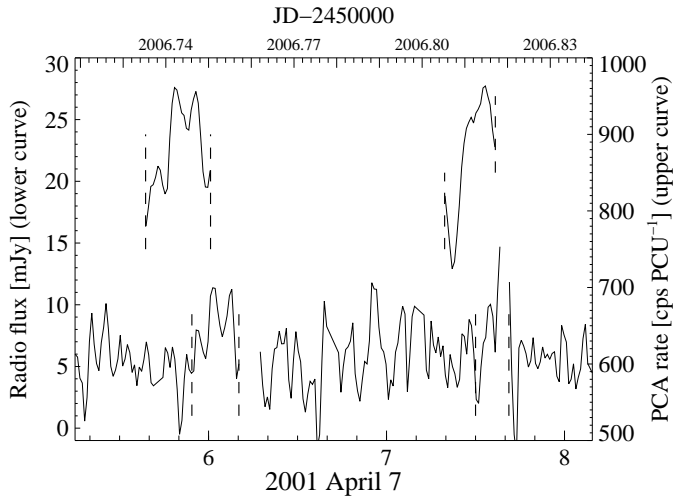}}
\caption{Cyg~X-1: Joint radio--X-ray lightcurves of 2001 April 7, binned to a
  resolution of 48\,s, showing a possible delay in the radio response
  to the X-ray variability (MCC is 0.75). Time on the $x$-axis is in
  hours.}
\label{fig:radio48_bothsg}
\end{figure}

Checking the simultaneous lightcurves of Cyg~X-1 by eye reveals several
observations that show a similar pattern in the X-ray and radio lightcurve,
with a delay in the radio response in the range of minutes.  The time
delays between X-ray and radio are not constant in all observations but
vary from a few minutes to tens of minutes.  Fig.~\ref{fig:radio48_bothsg}
shows the most convincing of these observations. The first X-ray lightcurve
segment leads the radio echo by $\sim$13\,min whereas the second X-ray
segment lies not more than $\sim$8\,min before the corresponding radio
pattern. In order to check whether such correlations are significant,
however, more statistical tests are required.

If a radio and an X-ray lightcurve are truly correlated, the corresponding
MCC will have an absolute value close to 1. Due to statistical reasons,
however, there is a finite probability of finding a large absolute
value of MCC even for two random lightcurves, e.g., white noise
lightcurves.  For a large sample of simultaneous radio and X-ray lightcurve
pairs, however, the distribution functions of MCCs computed for pairs with
random correlations and computed for pairs with a real correlation will be
different. By comparing the MCC distributions computed from white noise
lightcurves with those from the real data, it is thus possible to test for
the presence of a correlation between the radio and the X-rays in a
statistical sense.  In this section we show how such comparisons are
performed. 

As described by \citet{gleissner:04a}, we used a set of 120
simultaneous radio--X-ray lightcurve segments of GRS~1915+105
\citep{klein-wolt:02} to test the sensitivity of the cross-correlation
procedure used here. One peculiarity of the GRS~1915+105 lightcurves
is that they usually exhibit quasi-periodic dips in the X-ray emission
which are related to radio oscillations with a typical period
$P_\text{osc}$ of $\sim 20$--$45$\,\mbox{min}
\citep{pooley:97,mirabel:98a}. Assuming a one-to-one relation between
an X-ray dip and a subsequent oscillation peak in the radio emission,
\citet{klein-wolt:02} arrived at an estimated time delay of
14--30\,\mbox{min} of radio with respect to X-ray. This one-to-one
relation is equivalent to an anti-correlation with negative MCC close
to $-1$. When the shift of the X-ray lightcurve relative to the radio
lightcurve is continued for $P_\text{osc}/2$, i.e., where the X-ray
dip coincides with the subsequent oscillation valley of the radio
emission, we obtain a positive MCC close to $+1$. Since we use the
maximum value of the absolute value of the cross-correlation
coefficient as an indicator for a correlation, due to the noise in the
MCC these correlations appear as an anti-correlation in about half of
all cases, while in the remaining half it is the positive correlation
when the shift has been continued for $P_\text{osc}/2$.  Thus, both
negative and positive MCCs close to $\pm 1$ indicate the same
features, i.e., the radio--X-ray correlations of
\citet{klein-wolt:02}. We did not correct for this periodical effect
in the histograms of MCC, but accounted for it when determining the
corresponding time lag.

\begin{figure*}
\centering
   \includegraphics[width=17cm]{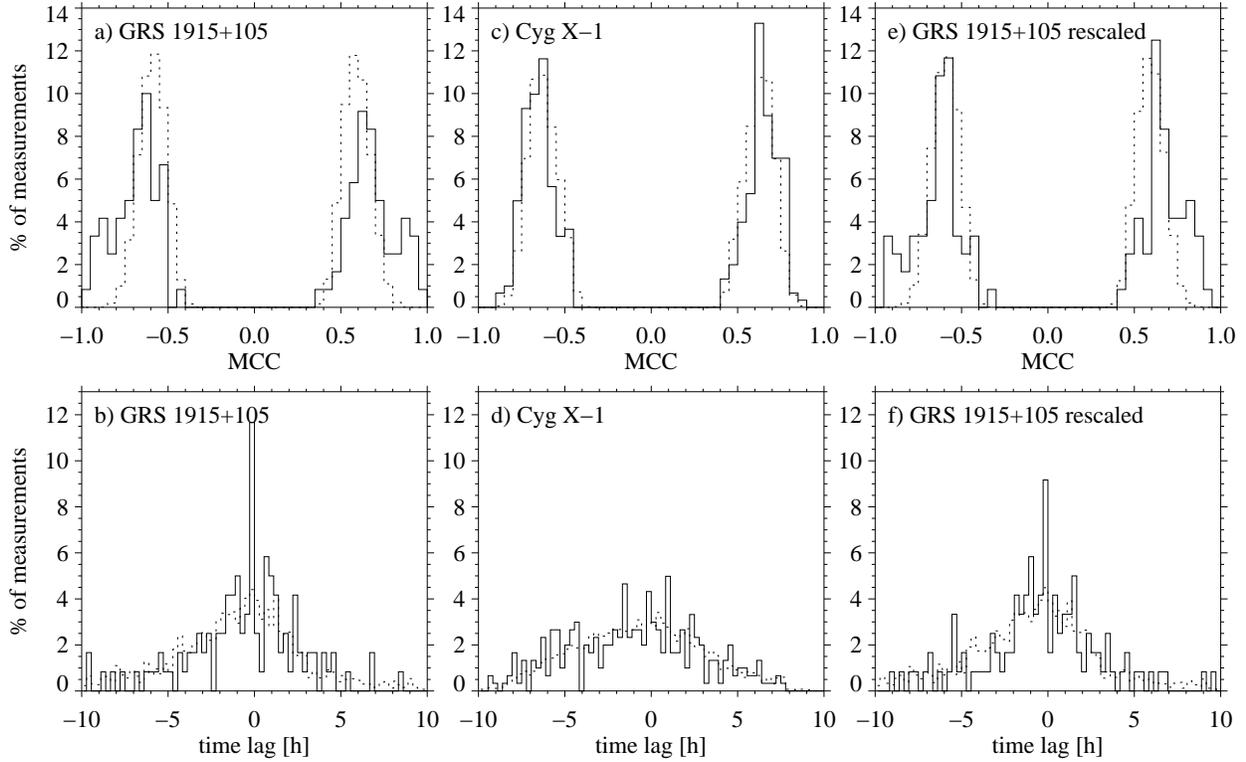}
\caption{\textbf{a)--b)} GRS~1915+105: Histograms of the distribution
  of MCC and the distribution of time lags for observed data (solid
  line) and simulated white noise data (dotted line), respectively.
  The histograms of GRS~1915+105 take into account 120 separate
  lightcurve pairs. The MCC have been binned to 40 bins with bin
  widths of 0.05 from $-1$ to $+1$, the time lags have been binned to
  72 bins with bin widths of 1000\,s from $-10$\,h to $+10$\,h.
  Original lightcurves have been binned to 32\,s and have been
  smoothed by the Savitzky-Golay filter. \textbf{c)--d)} The
  corresponding histograms for Cyg~X-1, taking into account 301
  separate lightcurve pairs. \textbf{e)--f)} The corresponding
  histograms for the GRS~1915+105 data, rescaled to the $S/N$-level of
  Cyg~X-1 (solid line) and simulated white noise data (dotted
  line).}
\label{fig:histogram_cyggrs}
\end{figure*}

Fig.~\ref{fig:histogram_cyggrs}a compares the histogram
$h_{\text{obs}}({\text{MCC}})$ of the MCC for the 120 simultaneous
radio and X-ray observations of GRS~1915+105 (solid line) with the
histogram $h_{\text{sim}}({\text{MCC}})$ of MCC from 1000 simulated
white noise data sets (dotted line). In this simulation we created
random white noise lightcurves for the radio and X-ray data with the
same mean value, standard deviation, and sampling as the observed
radio and X-ray lightcurves. Then we cross-correlated these simulated
lightcurves exactly the way we did with the observed ones and
determined the histogram of the corresponding MCC. For the whole data
set, i.e., the 120 simultaneous radio and X-ray observations, we ran
the simulation 1000 times in order to achieve a sufficient statistical
significance. Using a similar technique, we compared the histogram of
the observed time lags (determined from the MCC) with the distribution
of the time lags obtained from the Monte Carlo simulations
(Fig.~\ref{fig:histogram_cyggrs}b).

Figs.~\ref{fig:histogram_cyggrs}a and b show that both the MCC and the
time lag distribution for the GRS~1915+105 observations are
considerably different from the corresponding white noise
distributions. We use the Kolmogorov-Smirnov test statistic $D$
\citep{keeping:62} to quantify the difference of two distributions
(see Table~\ref{tab:ks-test}a). From Fig.~\ref{fig:histogram_cyggrs}a
we determine $D_\text{obs}$, giving us a measure of the difference
between the observed distribution of the data set and a standard white
noise distribution, obtained by averaging over 1000 simulated white
noise data sets.  The Kolmogorov-Smirnov test allows the determination
of the probability $P$ that the difference $D_\text{test}$ between a
single simulated white noise data set and the standard white noise
distribution is equal to or greater than $D_\text{obs}$. If $P$ is
sufficiently small, the null hypothesis $H_0$ that the observed and
the simulated distributions have been drawn from the same underlying
distribution function is to be rejected. Carrying out this analysis
for both the MCC and the lag distributions, we find that the null
hypothesis is to be rejected at the 0.1\% level for the MCC, and at
the 1.7\% level for the time lags (see Table~\ref{tab:ks-test}a).

\begin{table}
\caption{Characteristics of the Kolmogorov-Smirnov test for two
  samples: test statistic $D_\text{obs}$ for observed and standard white noise
  distribution, and probability $P$ for $D_\text{test} \ge D_\text{obs}$.}\label{tab:ks-test}   
\noindent
\begin{tabular}{lrrrr}
\hline
\hline
& \multicolumn{2}{l}{a) GRS~1915+105} & \multicolumn{2}{l}{b) Cygnus~X-1}\\
& MCC & time lag & MCC & time lag\\
\hline
$D_\text{obs}$   & 0.168 & 0.097  & 0.058 & 0.056 \\
$P [\%]$ & 0.1   & 1.7    & 13.6  & 14.4 \\
\end{tabular}
\end{table}

As expected, the radio--X-ray correlations in GRS~1915+105 which
were examined by \citet{klein-wolt:02} are reflected by the significant
fraction of MCCs with absolute values close to 1. This means that our
procedure is capable of finding radio--X-ray correlations in a data set.
Furthermore, the time lags that correspond to the MCC distribution for
GRS~1915+105 appear as a clearly increased number of observations with
time lag in the bin [$-1000$\,s,0\,s], consistent with earlier observations
\citep{klein-wolt:02}.

We now turn to the simultaneous \textsl{RXTE}--radio observations of
Cyg~X-1. Fig.~\ref{fig:histogram_cyggrs}c compares the histogram
$h_{\text{obs}}({\text{MCC}})$ of our 301 simultaneous radio and
\textsl{PCA} X-ray observations of Cyg~X-1 with
$h_{\text{sim}}({\text{MCC}})$ from 1000 white noise simulations of this
data set. The Kolmogorov-Smirnov test shows that the two histograms bear
significant similarities such that the null hypothesis of identical
distributions is to be accepted (Table~\ref{tab:ks-test}b). We conclude
that the MCC distribution of the observed and the simulated lightcurves are
similar. The same conclusion is drawn from a comparison of the
distributions of the time lags (Fig.~\ref{fig:histogram_cyggrs}d and
Table~\ref{tab:ks-test}b).  Unlike GRS~1915+105, the time lag distribution
of Cyg~X-1 does not show any distinctive features and is consistent with
that of the white noise simulation. This leads us to the assumption that
the similar patterns seen on short time scales in the X-ray and in the
radio lightcurves are random events which are a natural outcome in white
noise lightcurves.

Although more than 4 times closer to us, Cyg~X-1 is relatively dim in
X-ray and radio compared to GRS~1915+105. This raises the question:
is the negative result of Cyg~X-1 caused by the low signal to noise,
$S/N$, level, or would our correlation procedure be able to find
radio--X-ray correlations in Cyg~X-1 if they were present the same way
as in GRS~1915+105 but on the lower $S/N$ level of Cyg~X-1?

In order to answer this question we scaled the X-ray and radio lightcurves
of GRS~1915+105 to the level of Cyg~X-1, adding the appropriate
amount of Gaussian noise. Fig.~\ref{fig:histogram_cyggrs}e shows that the
MCC distribution of these rescaled lightcurves differs from that of the
corresponding white noise simulation distribution, similarly to
Fig.~\ref{fig:histogram_cyggrs}a.  Also, the time lag distribution of the
rescaled lightcurves (Fig.~\ref{fig:histogram_cyggrs}f) shows significant
differences with respect to the white noise simulation, similar to the
unscaled data of GRS~1915+105. This result proves that our correlation
procedure is sufficiently sensitive to detect any radio--X-ray correlations
in Cyg~X-1 if these are at the level of GRS~1915+105. We conclude that the
lack of radio--X-ray correlations on short time scales in Cyg~X-1 is not
caused by an insufficient $S/N$ ratio.

\subsection{Long term correlations}\label{sec:Long-term}

\begin{figure}
\resizebox{\hsize}{!}{\includegraphics{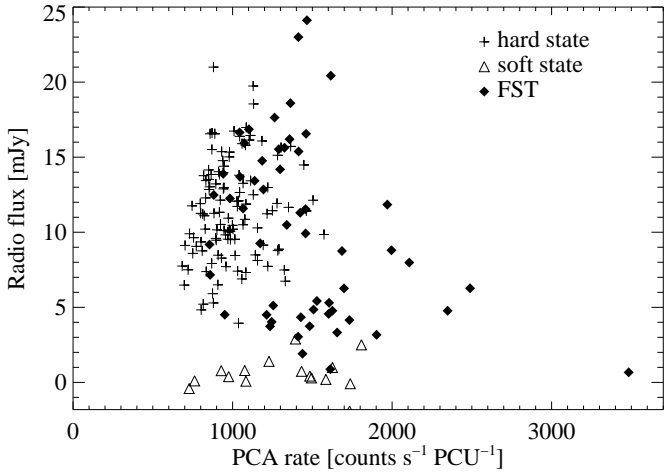}}
\caption{Cyg~X-1: \textsl{PCA} X-ray and radio flux-flux plot for data from
  1999 Jan 14 to 2003 Apr 06 with $0.25 \le \Phi_{\text{orb}} \le 0.75$,
  \textsl{PCA} channels 0--128 (mainly sensitive in the range
  $\sim$2--15\,keV).}
\label{fig:ryle_vs_pca1}
\end{figure}

\begin{figure}
\resizebox{\hsize}{!}{\includegraphics{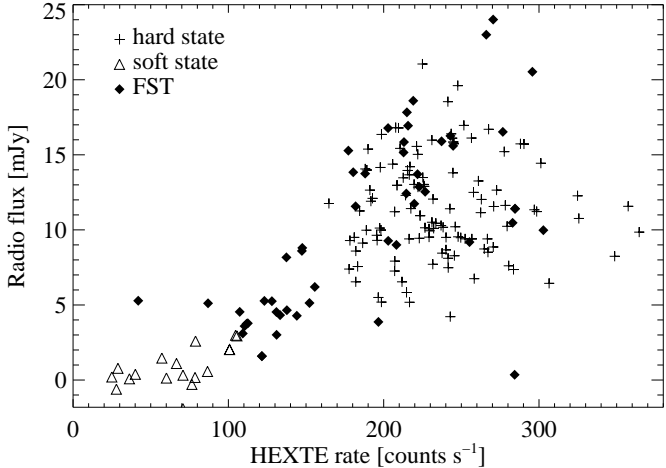}}
\caption{Cyg~X-1: \textsl{HEXTE} X-ray and radio flux-flux plot for data from
  1999 Jan 14 to 2003 Apr 06 with $0.25 \le \Phi_{\text{orb}} \le
  0.75$, \textsl{HEXTE} channels 15--255 ($\sim$15--255\,keV).}
\label{fig:ryle_vs_hexte1}
\end{figure}

As noted in Sect.~\ref{sec:intro}, no clear correlation pattern on long
term time scales between the radio emission and the \textsl{RXTE ASM} soft
(2--10\,keV) X-ray flux is found \citep{pooley:99a,brocksopp:99a}. The
availability of the pointed observations allows us to extend these studies
to harder energies, using the data from the \textsl{PCA} (which is mainly
sensitive below 15\,keV) and the \textsl{HEXTE} (channels 15--255,
sensitive above 15\,keV).  All observations from hard and soft states as
well as from FST have been included (for state definition see
Sect.~\ref{sec:data}). To account for known orbital modulation
\citep{brocksopp:99a,brocksopp:99b,brocksopp:02a}, we limit the data set in
orbital phase, $\Phi_\text{orb}$, to observations with $0.25 \le
\Phi_{\text{orb}} \le 0.75$.

Fig.~\ref{fig:ryle_vs_pca1} shows that no correlation between the soft
X-rays from the \textsl{PCA} data and the radio exists, confirming earlier
\textsl{ASM} results. We note that this pattern is similar to that seen in
GX~339$-$4 \citep{corbel:00a}. For the \textsl{HEXTE} band, above 15\,keV,
however, a long term correlation between the radio and the X-rays is
evident, particularly for the FST (Fig.~\ref{fig:ryle_vs_hexte1}).
Spearman rank correlation coefficients, $r_\text{S}$, have been calculated
for the subplots of hard state and FST data points, and for all data points
shown in Figs.~\ref{fig:ryle_vs_pca1} and \ref{fig:ryle_vs_hexte1},
confirming this result (see Table~\ref{tab:spearman}).

\begin{table}
\caption{Spearman rank correlation coefficients, $r_\text{S}$, of plots in
  Figs.~\ref{fig:ryle_vs_pca1} and \ref{fig:ryle_vs_hexte1}. The range of
  the correlation coefficient is $-1 \le r_\text{S} \le 1$, with a perfect
  correlation being indicated by $r_\text{S} = 1$ \citep{keeping:62}.}  
\label{tab:spearman}   
\noindent
\begin{tabular}{lccc}
\hline
\hline
& Hard state & FST & All data points \\
\hline
Radio/\textsl{PCA} & 0.17 & $-0.37$ & $-0.16$ \\
Radio/\textsl{HEXTE} & 0.05 & 0.67 & 0.50 \\
\end{tabular}
\end{table}

In terms of Comptonization models, the $>$15\,keV band is dominated by
Compton reflection and emission from the accretion disk corona, while an
appreciable fraction of the soft X-rays can be due to the accretion disk
\citep[e.g.,][]{dove:97a}. The correlation displayed in
Fig.~\ref{fig:ryle_vs_hexte1} shows that the radio emission is directly
linked to this hard component of the X-ray spectrum, which is present
during the state transitions and the hard state, but not during the soft
state, consistent with the earlier findings cited in Sect.~\ref{sec:intro}.
The correlation is also remarkably similar to the correlation between the
rms variability at $>$15\,keV and the \textsl{BATSE} hard X-ray flux
discovered by \citet{crary:96}.

\section{Summary and Discussion}\label{sec:discussion}

In this paper we have described the correlation between the radio and
the X-ray emission of Cyg~X-1 as seen with observations with the Ryle
Telescope and the pointed instruments on \textsl{RXTE}. The major
result of our analysis is that there is a correlation between the
radio flux and the hard X-rays, particularly during intermediate
states (flares, transitions, etc.), on time scales of days and weeks
to months (Fig.~\ref{fig:ryle_vs_hexte1}), and no clear correlation
between the soft X-rays and the radio flux on this time scale
(Fig.~\ref{fig:ryle_vs_pca1}).  This result lends further support to
the jet/disk concept \citep{brocksopp:99a,fender:02a,markoff:03a}.
This overall connection of disk and jet has been affirmed in many
observations, while the geometry at the base of the jet is still
unclear. During the soft state, the radio emission is quenched and no
jet is produced. In the hard state, the outflow from the accretion
disk into the jet is relatively smooth, as represented by continuous
radio emission over the length of the jet \citep{stirling:01a},
washing out smaller variability features. When the mass accretion rate
in the disk varies significantly on time scales of several hours to
days, this behavior is seen as X-ray flares and corresponding radio
flares, resulting in radio--X-ray correlations during the flaring
state.

On time scales of minutes to $\sim$5\,hours, no statistically significant
correlation could be detected (Fig.~\ref{fig:histogram_cyggrs}). An
explanation for this behavior could be that on time scales shorter than
the propagation time between emission regions other factors, e.g.,
intervening turbulence, smooth out or distort any variability present.

\begin{figure}
\resizebox{\hsize}{!}{\includegraphics{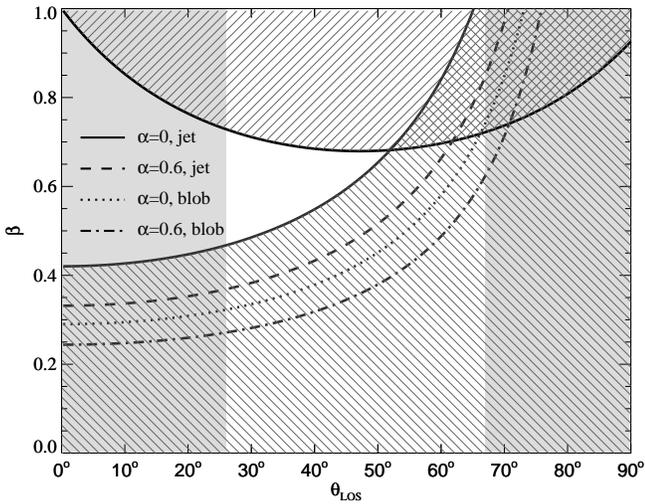}}
\caption{Limit on jet speed $\beta=v/c$ as a function of the angle
  between a jet and our line of sight, $\theta_\text{LOS}$. Upper
  hatched region: exclusion region from the nondetection of correlated
  radio--X-ray variability. Lower hatched region: excluded region of
  the parameter space from jet imaging \citep{stirling:01a}, for a
  continuous and a clumpy jet, and radio spectral index $\alpha=0$ and
  $\alpha=0.6$. The grey regions are the current limits on the
  inclination of the orbit of HDE~226868, and assume that the jet is
  perpendicular to the orbital plane.}\label{fig:betlim}
\end{figure}

Using the nondetection of correlated variability allows us to set limits on
the speed $\beta=v/c$ of the jet and on its inclination.  The time for
traversing a free jet (with no additional acceleration) is given by
$t=d/\beta c$ where $d$ is the distance where the 15\,GHz emission is
produced and where $\beta$ is the speed of the jet.  Simultaneous broadband
jet model fits \citep{markoff:03a} as well as VLBA observations at 8.4\,GHz
\citep{stirling:01a} place $d$ in the range of $d\sim
10^{12}$--$10^{14}$\,cm. The jet speed is found to be $\beta\sim
0.2$--$0.3$ from broadband analyses \citep{gallo:03a,maccarone:03a}, with
an upper limit of $\beta\sim 1$ from ballistic ejection events in MQs
(e.g., $\Gamma \sim 17$, $\beta \sim 1$, in V4641~Sgr, M.~Rupen, priv.\ 
comm.). The travel time expected from these numbers is between
$10^{12}\,{\text{cm}}/c \approx 33\,\text{s}$ and
$10^{14}\,{\text{cm}}/0.2c \approx 4.6\,\text{h}$.  Depending on the
inclination of the jet, the delay between the X-ray and the radio signals
when they reach the observer will be in the same time range.  Similar
constraints are obtained from VLBA imaging, where we find that for the most
probable case of a continuous jet with flat spectrum \citep{fender:00a} and
an angle of the jet with respect to our line of sight, $\theta_{\text{LOS}}
\approx$ 35\hbox{$^\circ$}--\,40\hbox{$^\circ$} \citep{herrero:95},
$\Delta\tau_\text{max}$ amounts to $2\cdot 4.2\,\text{h} \approx
8\,\text{h}$. Assuming that at 15\,GHz the radio jet of Cyg~X-1 has at most
the same extent as the 8.7\,GHz jet seen by VLBA, Fig.~\ref{fig:betlim}
shows that the nondetection of correlated radio--X-ray variability at
timescales under 5\,hours limits the jet speed to $\beta=0.7$ for a
continuous jet with a flat radio spectrum, with a lower limit of $\beta\sim
0.4$, depending on $\theta_\text{LOS}$ and consistent with earlier results
\citep{stirling:01a}. In the likely case that the 15\,GHz jet is smaller
than the 8.7\,GHz jet, these limits are even tighter.

We conclude that given that no radio--X-ray correlations on
minute to hour time scales can be identified and given that
correlations on day to month time scales are observed, it is
consequent to look for correlations on intermediate time scales like
hours to days. Probing these intermediate time scales will be the aim
of an approved \textsl{RXTE} observation during 2004.

\begin{acknowledgements}
  T.G.\ is supported by the Deutsche Forschungs\-gemeinschaft through
  grant Sta\,173/25. S.M.\ is supported by an National Science
  Foundation (NSF) Astronomy \& Astrophysics Postdoctoral Fellowship,
  under award AST-0201597. S.H.\ is supported by the National
  Aeronautics and Space Administration through \textsl{Chandra}
  Postdoctoral Fellowship Award Number PF3-40026 issued by the
  \textsl{Chandra} X-ray Observatory Center, which is operated by the
  Smithsonian Astrophysical Observatory for and on behalf of the
  National Aeronautics Space Administration under contract NAS8-39073.
  The Ryle Telescope is supported by PPARC. This research has made use
  of data obtained from the High Energy Astrophysics Science Archive
  Research Center, provided by NASA's Goddard Space Flight Center. We
  acknowledge travel funds from the Deutscher Akademischer
  Austauschdienst and the NSF.
\end{acknowledgements}

\end{document}